\documentclass[prl,letterpaper,twocolumn,showpacs,superscriptaddress,amsmath,amsfonts,amssymb,preprintnumbers]{revtex4}
\pdfoutput=1
\usepackage[T1]{fontenc}\usepackage[latin1]{inputenc}
\usepackage{dcolumn,graphicx,color,hvfloat,booktabs,microtype,afterpage} \graphicspath{{Figures/}}
\usepackage[charter]{mathdesign}
\renewcommand{\figurename}{Fig.}
\renewcommand{\tablename}{Table}
\makeatletter\renewcommand{\fnum@figure}[1]{\figurename~\thefigure.}\makeatother
\makeatletter\renewcommand{\fnum@table}[1]{\tablename~\thetable.}\makeatother

\usepackage[colorlinks,plainpages=false,linkcolor=blue,urlcolor=blue,citecolor=black,pdfpagemode=UseNone,pdfstartview=FitBH]{hyperref}

\newcommand{\rfs }{\mbox{Rb$_2$Fe$_4$Se$_5$}}

\begin{document}

\title{Magnetic Resonant Mode in the Low-Energy Spin-Excitation Spectrum\\ of Superconducting \rfs\ Single Crystals}

\author{J.\,T.~Park}\author{G.\,Friemel}\author{Yuan\,Li}\author{J.-H.\,Kim}
\affiliation{Max-Planck-Institut für Festkörperforschung, Heisenbergstraße 1, 70569 Stuttgart, Germany}

\author{V.~Tsurkan}
\affiliation{\mbox{Experimentalphysik V, Center for Electronic Correlations and Magnetism, Institute for Physics, Augsburg Univ., 86135 Augsburg, Germany}}
\affiliation{Institute of Applied Physics, Academy of Sciences of Moldova, MD 2028, Chisinau, Republic of Moldova}

\author{J.\,Deisenhofer}\author{H.-A.\,Krug\,von\,Nidda}\author{A.\,Loidl}
\affiliation{\mbox{Experimentalphysik V, Center for Electronic Correlations and Magnetism, Institute for Physics, Augsburg Univ., 86135 Augsburg, Germany}}

\author{A.\,Ivanov}
\affiliation{Institut Laue-Langevin, 6 rue Jules Horowitz, 38042 Grenoble Cedex 9, France}

\author{B.\,Keimer}\author{D.\,S.\,Inosov{\large\hyperref[CorrAuthor]{*}}}
\affiliation{Max-Planck-Institut für Festkörperforschung, Heisenbergstraße 1, 70569 Stuttgart, Germany}

\begin{abstract}
We have studied the low-energy spin-excitation spectrum of the single-crystalline \rfs\ superconductor ($T_{\rm c}=32$\,K) by means of inelastic neutron scattering. In the superconducting state, we observe a magnetic resonant mode centered at an energy of $\hslash\omega_{\rm res} = 14$\,meV and at the $(0.5~0.25~0.5)$ wave vector (unfolded Fe-sublattice notation), which differs from the ones characterizing magnetic resonant modes in other iron-based superconductors. Our finding suggests that the 245-iron-selenides are unconventional superconductors with a sign-changing order parameter, in which bulk superconductivity coexists with the $\sqrt{5} \times \sqrt{5}$ magnetic superstructure. The estimated ratios of $\hslash\omega_{\rm res}/k_{\rm B}T_{\rm c} \approx 5.1 \pm 0.4$ and $\hslash\omega_{\rm res}/2\Delta \approx 0.7 \pm 0.1$, where $\Delta$ is the superconducting gap, indicate moderate pairing strength in this compound, similar to that in optimally doped 1111- and 122-pnictides.
\end{abstract}


\maketitle
Soon after the discovery of arsenic-free iron-selenide superconductors $A_{2}{\rm Fe_4}{\rm Se_5}$ ($A =$ K, Rb, Cs), also known as 245-compounds \cite{Discovery}, their unprecedented physical properties came to light, such as the coexistence of high-$T_{\rm c}$ superconductivity with strong antiferromagnetism \cite{Dispute,Chen11}. The pairing mechanism and the symmetry of the superconducting (SC) order parameter in this family of compounds still remain among the major open questions. In the majority of other Fe-based superconductors, it is widely accepted that the strong nesting between the holelike Fermi surface (FS) at the Brilliouin zone (BZ) center and electronlike FS at the BZ boundary leads to the sign-changing $s$-wave ($s_\pm$-wave) pairing symmetry \cite{Mazin08}. This scenario has been supported by different experimental methods, such as angle-resolved photoemission spectroscopy (ARPES) \cite{ARPES}, quasi-particle interference \cite{Hanaguri10}, and inelastic neutron scattering (INS) \cite{Resonance,Shamoto10}. On the other hand, recent theoretical calculations \cite{Bandstructure} and ARPES experiments \cite{Qian11,Gap} on the 245-system revealed the absence of holelike FS at the BZ center in the electronic structure, implying that the nesting between the hole- and electronlike FS sheets is no longer present. Hence, several theoretical studies proposed alternative pairing instabilities, such as $d$-wave or another type of $s_\pm$-wave symmetry with sign-changing order parameter between bonding and anti-bonding states \cite{Maier11,Pairing1,Pairing2}. As a hallmark of sign-changing SC order parameter, several authors theoretically predicted a resonant mode in the magnetic excitation spectrum below the SC transition, yet its precise position in momentum space still remains controversial \cite{Maier11,Pairing1}.

\begin{figure}[b]
\vspace{-1em}
\includegraphics[width=0.85\columnwidth]{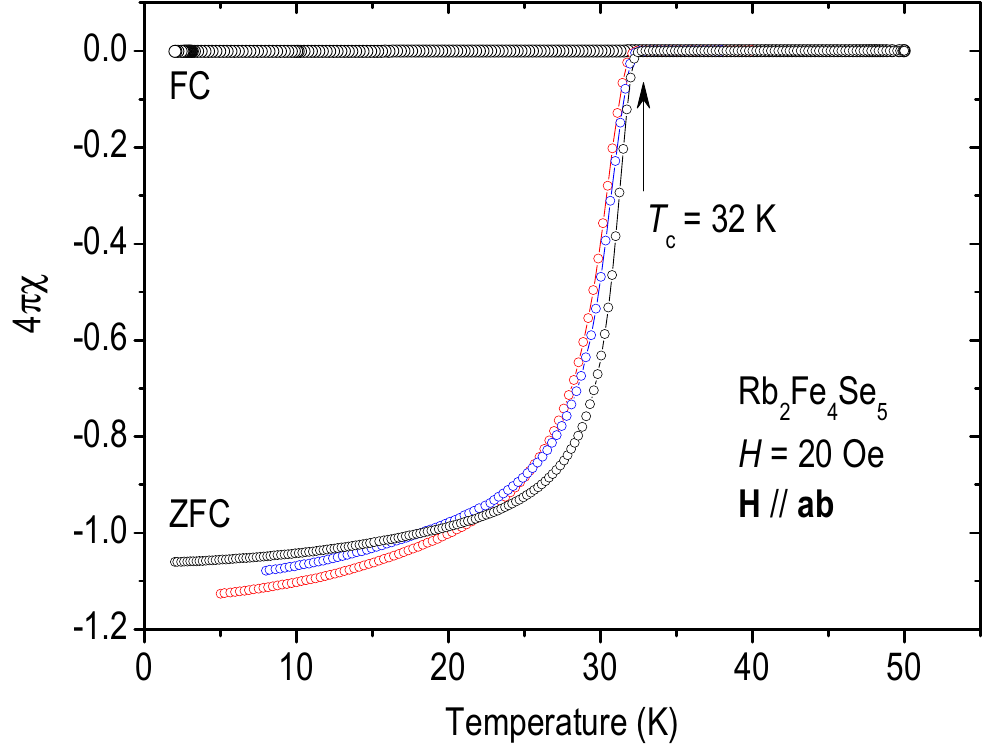}
\vspace{-1em}
\caption{The dc magnetic susceptibility measurements on three representative single-crystalline \rfs\ samples from the same batch. A sharp diamagnetic response is observed in the ZFC measurement right below 32\,K, indicating $\sim$100\% exclusion of the external magnetic field.}
\label{fig:characterization}
\vspace{-1em}
\end{figure}

A major complication in treating the 245-compounds theoretically arises from the presence of a crystallographic superstructure of Fe vacancies that has been consistently reported both from x-ray and neutron diffraction experiments \cite{Diffraction}. This $\sqrt{5}\times\sqrt{5}$ superstructure is closely related to the static antiferromagnetic (AFM) order persisting up to the N\'{e}el temperature, $T_{\rm N} \approx 540$\,K \cite{Yan11}. Although most of the existing band structure calculations have so far neglected the superstructure, several others have pointed out that it may have a strong influence on the FS shape \cite{MagneticFS,FSrecons}. However, these pronounced FS reconstruction effects have not been experimentally confirmed so far \cite{Chen11,Qian11,Gap}. Such an uncertainty in the FS geometry and its nesting properties makes it hard to predict the exact location of itinerant spin fluctuations in reciprocal space. Moreover, band structure calculations in the vacancy-ordered magnetic state result in insulating solutions for the stoichiometric 245-compound \cite{MagneticFS}. A possible way to reconcile these apparently contradictory observations is to assume a nanoscale phase separation of (i) insulating vacancy-ordered magnetic domains and (ii) metallic non-magnetic phase domains with effective electron doping that could host superconductivity at low temperature. Such kind of electronic phase segregation, resembling the situation in hole-doped 122-pnictides \cite{Park09}, found recent support from ARPES \cite{Chen11}, scanning nano-focus single-crystal x-ray diffraction imaging \cite{Ricci11}, scanning-tunneling microscopy \cite{Li11}, and optical spectroscopy \cite{Charnuka11}. In this letter, we provide experimental insight by using INS to directly probe the elementary magnetic excitations in superconducting Rb$_2$Fe$_4$Se$_5$ (RFS).

\begin{figure}[t]
\includegraphics[width=1.0\columnwidth]{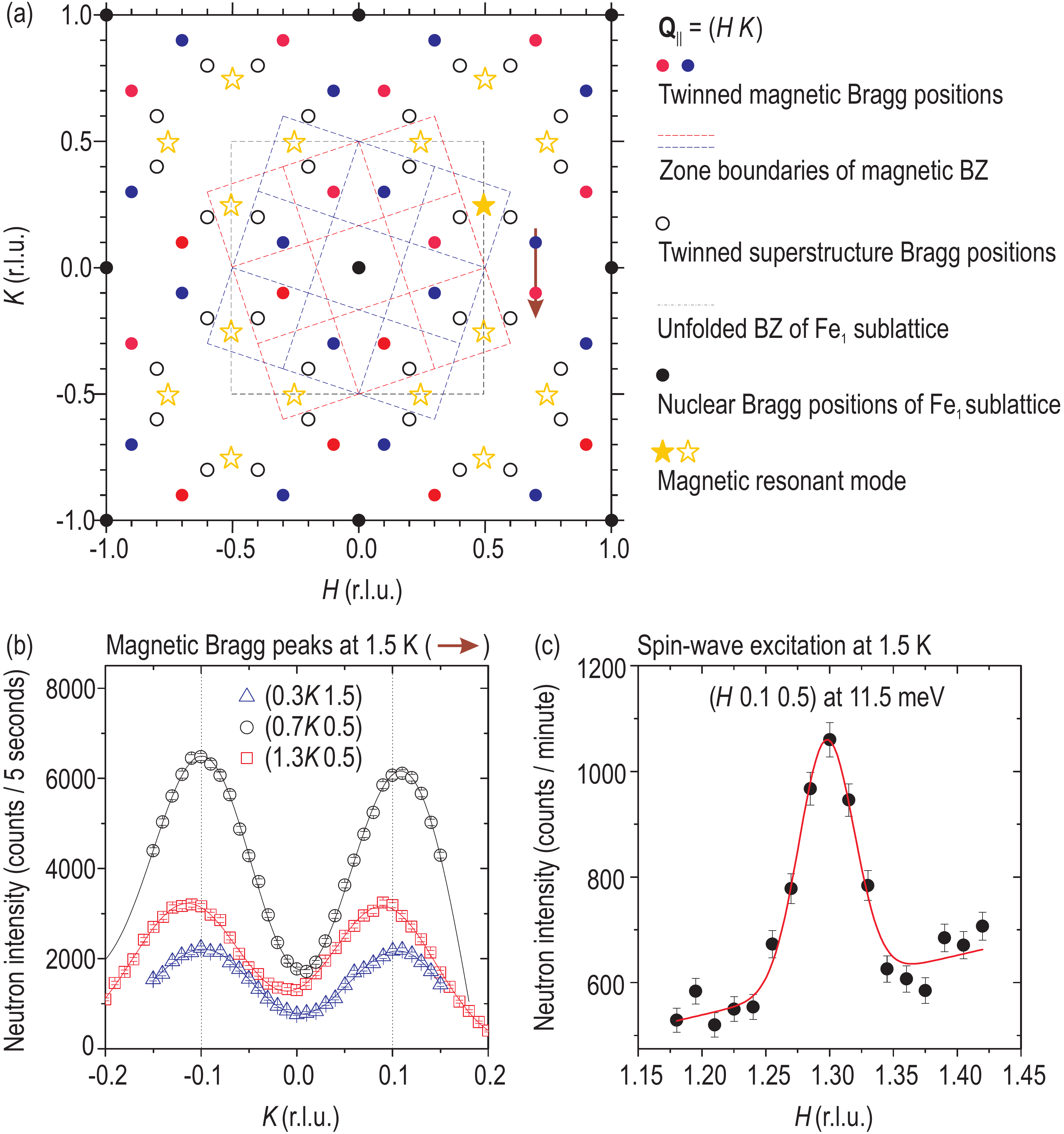}
\caption{(a)~The in-plane projection of twinned magnetic and nuclear Bragg peak positions arising from the $\sqrt{5} \times \sqrt{5}$ Fe-vacancy superstructure. The two sets of dots and the corresponding dashed lines (red and blue) denote magnetic superstructure Bragg reflections and the magnetic BZ boundaries for the right and left twin domains, respectively. Black dotted lines represent the Fe$_1$ unfolded BZ boundary. The arrow shows the trajectory of elastic momentum scans. (b)~Elastic scans along the $K$ direction [arrow in panel (a) and equivalent scans] measured at 1.5\,K. The almost identical neutron intensities of two symmetric magnetic Bragg peaks indicate the nearly equal population of the two twin domains in the sample. (c)~INS intensity at 11.5\,meV in the vicinity of the magnetic ordering wave vector $(1.3~0.1~0.5)_{\rm Fe_1}$. The intense spin-wave excitation peak is consistent with recent time-of-flight INS measurements on an insulating Rb$_{2+\delta}$Fe$_4$Se$_5$ compound \cite{Wang11}.}
\label{fig:magnetism}
\vspace{-2em}
\end{figure}

For the present study, we used a mosaic of RFS single crystals with a total mass of $\sim$\,1\,g, grown by the Bridgman method \cite{Tsurkan11}. The nearly stoichiometric and homogeneous composition with Rb:Fe:Se = 0.796:1.596:2.000 (1.99:3.99:5) has been determined by wave-length dispersive x-ray electron-probe microanalysis using a \textit{Camebax SX50} analyzer with an accuracy of 0.5\% for Fe and up to 1\% for Se. The SC properties of the sample were characterized by magnetometry, where $\sim$100\% flux exclusion was observed in the zero-field-cooled (ZFC) measurement for temperatures up to $T_{\rm c} = 32$\,K  [Fig.\,1(a)]. The INS experiment was performed at the thermal-neutron triple-axis spectrometer IN8 (ILL, Grenoble), with the sample mosaic mounted in the $(HH0)/(00L)$ or $(H00)/(00L)$ scattering planes. The wave vectors $\mathbf{Q}=(0.5~0.5~L)_{\rm Fe_1}$ and $(0.5~0~L)_{\rm Fe_1}$ were directly accessible in our scattering planes, and the spectrometer further allowed us to tilt the sample in order to access $\mathbf{Q}$-vectors in a certain range out of the scattering planes. Here and henceforth, we are using unfolded reciprocal-space notation corresponding to the Fe sublattice, which we denote as Fe$_1$, because of its simplicity and the natural correspondence to the symmetry of the observed signal \cite{Park10}. We quote the wave vector $\mathbf{Q}=(HKL)$ in reciprocal-lattice units (r.l.u.), i.e. in units of the reciprocal-lattice vectors $\mathbf{a^*}$, $\mathbf{b^*}$, $\mathbf{c^*}$ of the Fe sublattice ($a^* = 2\pi/a$, etc.). Here $a=b=2.76$\,{\AA} is the room-temperature distance between the nearest-neighbor Fe atoms, and $c=7.25$\,{\AA} is the distance between Fe layers. All INS measurements were done in the fixed-$k_{\rm f}$ $(k_{\rm f}=2.662\,${\AA}$^{-1}$) mode, using double-focused PG(002) monochromator and analyzer. A 5\,cm thick oriented PG filter was
installed before the analyzer to eliminate higher-order neutron contamination, and no collimation was applied to maximize the intensity.

We start with the magnetic Bragg peak patterns arising from the $\sqrt{5} \times \sqrt{5}$ Fe-vacancy superstructure. Panel (a) in Fig.\,\ref{fig:magnetism}\ is a sketch of magnetic and nuclear superstructure Bragg reflections, projected on to the two-dimensional $\mathbf{Q}_{\parallel}=(H~K)$ plane. Black dots and the large dashed rectangle correspond to the center and boundaries of the unfolded Fe$_1$ BZ, respectively. The solid dots represent magnetic Bragg reflections from two twin domains, and the corresponding dashed lines, rotated clockwise and counterclockwise with respect to the Fe$_1$ BZ, are magnetic zone boundaries of the two twin domains. Forbidden magnetic Bragg peaks, which coincide with the nuclear Fe-vacancy superstructure reflections seen by x-ray diffraction \cite{Diffraction}, are shown by empty circles. Figure\,\ref{fig:magnetism}\,(b) shows elastic scans crossing two magnetic Bragg peaks at $(0.7~\pm0.1~0.5)_{\rm Fe_1}$, as shown by the arrow in panel (a), and along equivalent reciprocal-space directions in higher BZs. Note that the two magnetic Bragg peaks at $K=\pm0.1$ originate from different twin domains, so that their similar intensity indicates almost equal population of both twins in our sample. The magnetic Bragg peak intensity decreases more rapidly when moving to a higher BZ along the out-of-plane direction than in-plane, indicating that the magnetic moment is oriented predominantly along the $L$-direction in this system. This is consistent with the reported spin configuration in the magnetically ordered phase, in which spins are alternatively pointing up and down along the $c$-axis \cite{Ye11}. Figure\,\ref{fig:magnetism}\,(c) shows inelastic magnetic intensity in the vicinity of the AFM ordering wave vector $\mathbf{Q}=(1.3~0.1~0.5)_{\rm Fe_1}$ at 11.5~meV, measured at low temperature, $T=1.5$\,K. The intense spin-wave peak is consistent with recent time-of-flight INS measurements on an insulating Rb$_{2+\delta}$Fe$_4$Se$_5$ compound \cite{Wang11}.

\begin{figure}[t]
\includegraphics[width=1\columnwidth]{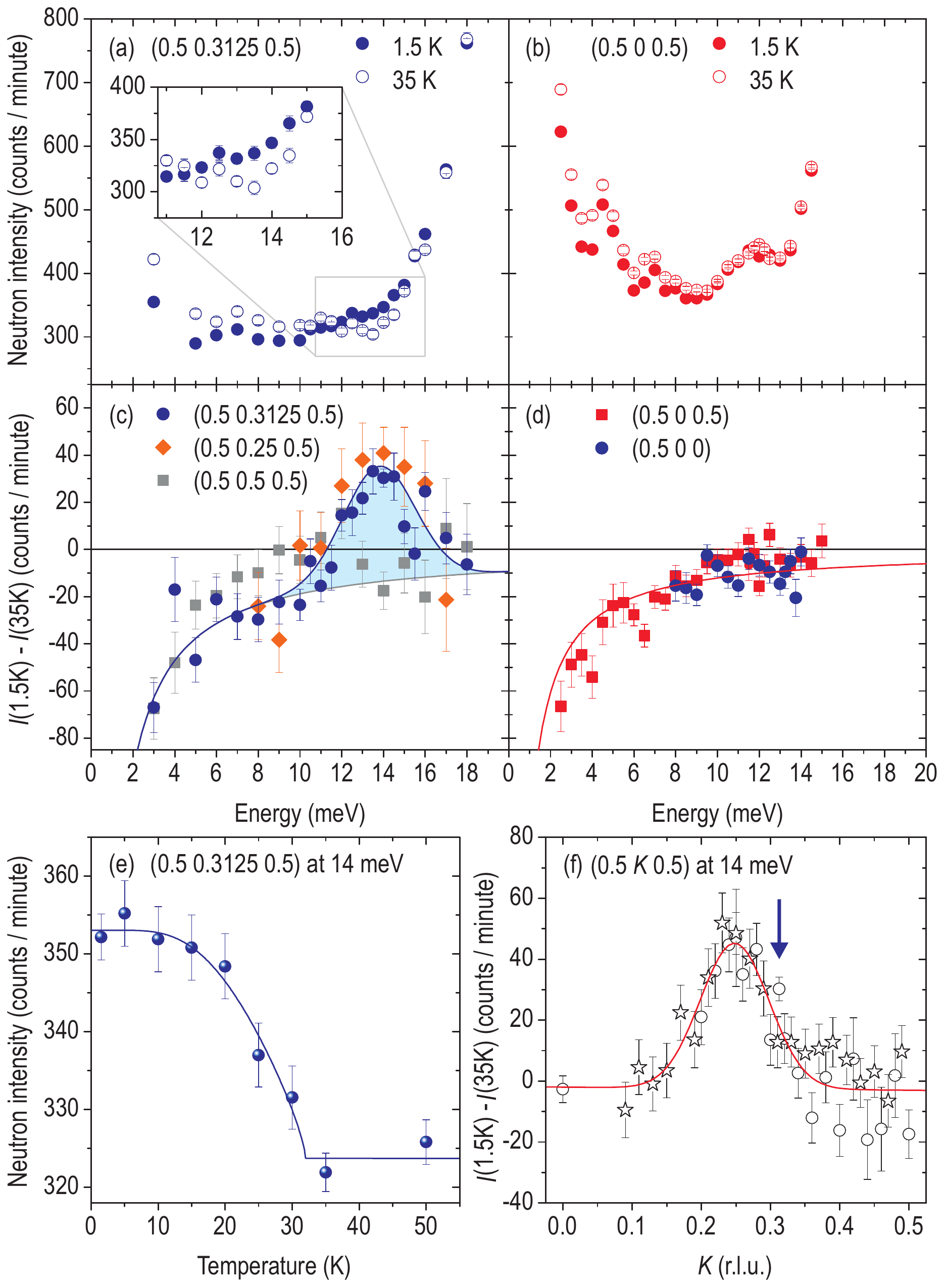}
\caption{(a,\,b)~Raw energy scans measured in the SC (1.5\,K) and normal (35\,K) states at $\mathbf{Q} = (0.5~0.3125~0.5)_{\rm Fe_1}$ and $(0.5~0~0.5)_{\rm Fe_1}$, respectively. The inset in panel (a) shows the zoomed-in part of the resonant peak in the raw data. (c)~Intensity difference between the SC state and the normal state at three $\mathbf{Q}$-vectors: $(0.5~0.25~ 0.5)_{\rm Fe_1}$, $(0.5~0.3125~0.5)_{\rm Fe_1}$, and $(0.5~0.5~0.5)_{\rm Fe_1}$. While there is no positive intensity at $(0.5~0.5~0.5)$, a clear resonance peak (shaded region) is observed around 14~meV both at $(0.5~0.25~0.5)_{\rm Fe_1}$ and $(0.5~0.3125~0.5)_{\rm Fe_1}$. (d)~The same plot as in panel (c), but for $\mathbf{Q} = (0.5~0~0.5)_{\rm Fe_1}$ and $(0.5~0~0)_{\rm Fe_1}$, where the magnetic resonant mode has been found in other Fe-based superconductors, but is absent here. The base line in (c) and (d) is the difference of the Bose factors. (e) Temperature dependence of the raw INS intensity at 14 meV and $\mathbf{Q} = (0.5~0.3125~0.5)_{\rm Fe_1}$ that demonstrates an order-parameter-like behavior with an onset at $T_{\rm c}$. (f)~Intensity difference of momentum scans along the BZ boundary, measured below and above $T_{\rm c}$, with a maximum at the commensurate wave vector $\mathbf{Q}_{\rm res} = (0.5~0.25~0.5)_{\rm Fe_1}$. The solid line is a Gaussian fit with a linear background. Different symbols represent identical momentum scans measured in different experiments, rescaled to the (002) nuclear Bragg peak intensity. The position of the resonant mode predicted by Maier \textit{et al.} \cite{Maier11} is shown by the arrow.}
\label{fig:resonance}
\vspace{-2em}
\end{figure}

Now we turn to the INS measurements across $T_{\rm c}$ near a few candidate $\mathbf{Q}$-vectors, where the magnetic resonant mode could be expected. Figure \ref{fig:resonance}, (a) and (b), displays raw energy-scan spectra recorded above and below $T_{\rm c}$ at $\mathbf{Q}=(0.5~0.3125~0.5)_{\rm Fe_1}$, where the resonance has been theoretically predicted \cite{Maier11}, and at $(0.5~0~0.5)_{\rm Fe_1}$, where it is usually found in other Fe-based superconductors \cite{Resonance,Shamoto10}. In the absence of any resonant enhancement, the intensity is expected to be higher in the normal state due to the influence of the Bose factor at low energies. Already in the raw data, one can see that this is the case for all data points except a narrow energy region around 14\,meV at $\mathbf{Q}=(0.5~0.3125~0.5)_{\rm Fe_1}$.

To emphasize this effect and to eliminate the energy-dependent background, we plot temperature differences of the same datasets in Figs.\,\ref{fig:resonance}\,(c) and (d). Also shown are the difference spectra for $\mathbf{Q}=(0.5~0.5~0.5)_{\rm Fe_1}$, $(0.5~0.25~0.5)_{\rm Fe_1}$, and $(0.5~0~0)_{\rm Fe_1}$. As seen in Fig.\,\ref{fig:resonance}\,(c), a prominent peak (shaded region) is found at $\hslash\omega_{\rm res}\approx14$\,meV for $\mathbf{Q}=(0.5~0.3125~0.5)_{\rm Fe_1}$ and $\mathbf{Q}=(0.5~0.25~0.5)_{\rm Fe_1}$, which we attribute to the magnetic resonant mode. However, no such peak is observed at $\mathbf{Q}=(0.5~0.5~0.5)_{\rm Fe_1}$, in contrast to some alternative predictions based on the $d$-wave pairing symmetry \cite{Pairing1}. Figure\,\ref{fig:resonance}\,(d) also demonstrates the absence of any resonant mode at $\mathbf{Q}=(0.5~0~0.5)_{\rm Fe_1}$ and $(0.5~0~0)_{\rm Fe_1}$, where it is usually found in iron pnictides \cite{Resonance,Shamoto10}. At these wave vectors, the data simply follow the solid line, which is the Bose-factor difference between 1.5\,K and 35\,K.

To verify whether the observed redistribution of spectral weight at low temperatures is related to the SC transition, we have measured the temperature dependence of the resonance intensity at $\mathbf{Q}=(0.5~0.3125~0.5)_{\rm Fe_1}$, which is shown in Fig.\,\ref{fig:resonance}\,(e). Indeed, an order-parameter-like increase of intensity below $T_{\rm c}$ is found, which is accepted as the hallmark of the magnetic resonant mode.

\begin{figure}[b]
\includegraphics[width=0.85\columnwidth]{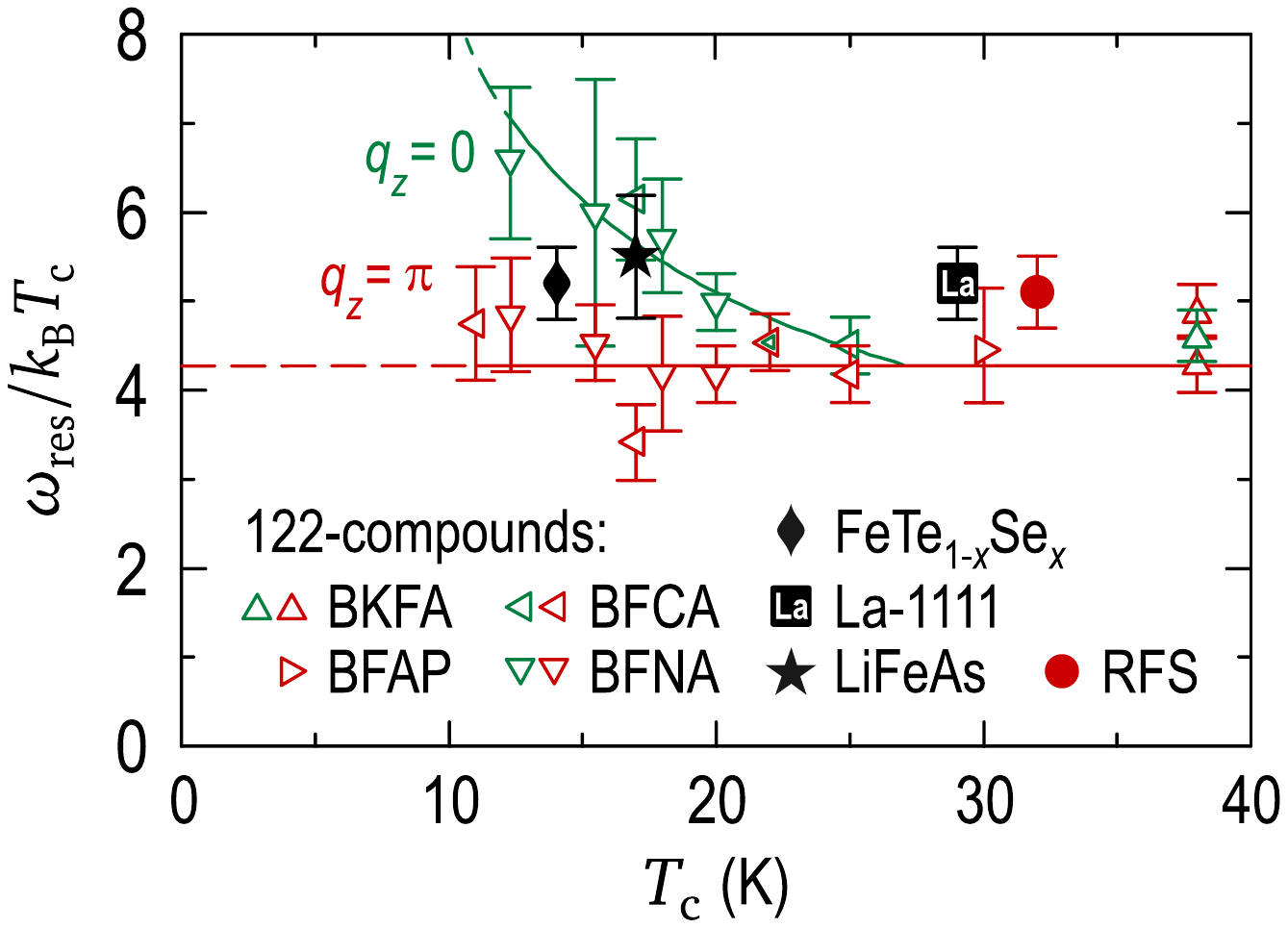}
\caption{Normalized resonance energy, $\hslash\omega_{\rm res}/k_{\rm B}T_{\rm c}$, in Fe-based superconductors for $q_{z}=0$ and $q_{z}=\pi$ \cite{Inosov11}. This ratio for RFS is slightly higher than for 122-compounds, but comparable to 11-, 111-, and 1111-type superconductors.}
\label{fig:gapres}
\end{figure}

To pin down the exact location of the resonance in $\mathbf{Q}$-space, we have measured momentum scans along the BZ boundary at both temperatures. Their difference is presented in Fig.\,\ref{fig:resonance}\,(f) and suggests a maximum at the commensurate nesting wave vector $\mathbf{Q}_{\rm res}=(0.5~0.25~0.5)_{\rm Fe_1}$ shown by the star symbols in Fig.\,\ref{fig:magnetism}\,(a), close to the predicted resonance position, $\mathbf{Q}=(0.5~0.3125~0.5)_{\rm Fe_1}$ \cite{Maier11}. Yet, the disagreement is small compared to the $\mathbf{Q}$ width of the peak, which explains the similar INS response at both $\mathbf{Q}$-vectors, as seen from Fig.\,\ref{fig:resonance}\,(c). Because the position of the nesting vector is strongly doping dependent, and the calculations in Ref.\,\onlinecite{Maier11} were done for the arbitrary doping of 0.1 electrons per Fe, a quantitative agreement with our results is not expected. The observed wave vector suggests an even higher level of the effective electron doping in the metallic phase of the sample, which is difficult to reconcile with the stoichiometric chemical composition unless we assume electronic phase segregation into electron-rich and electron-poor regions of the kind discussed in Ref.~\onlinecite{Ricci11} and \onlinecite{Li11}. Such a high doping level of the metallic regions would also agree qualitatively with the ARPES results \cite{Chen11,Qian11,Gap}.

By comparing the normalized resonance intensity in RFS with that in the nearly optimally doped Ba(Fe$_{1-x}$Ni$_x$)$_2$As$_2$ (BFNA), measured in a similar experimental configuration at the same spectrometer \cite{Park10}, we find that the intensity at the resonance energy in RFS is approximately a factor of three smaller than in BFNA. Because we expect four nonequivalent resonant peaks in the BZ of RFS from symmetry considerations, as opposed to only two such peaks [$\mathbf{Q}_{\rm res}=(0.5~0)_{\rm Fe_1}$ and $(0~0.5)_{\rm Fe_1}$] in the 122-system, the total resonant spectral weight in both compounds turns out to be comparable.

It has been shown that the resonance energy scales linearly with $T_{\rm c}$ in iron-based superconductors, with a ratio of $\hslash\omega_{\rm res}/k_{\rm B}T_{\rm c}$ that slightly varies between different families \cite{Shamoto10,Park10,Wang10}, but is generally lower than the respective ratio of $\sim$\,5.3 measured in high-$T_{\rm c}$ cuprates \cite{Eschrig06}. In Fig.\,\ref{fig:gapres}, we compare this ratio in all Fe-based superconductors, in which the resonant mode has been found [see Ref.~\citenum{Inosov11} and references therein]. The value for RFS, $\hslash\omega_{\rm res}/k_{\rm B}T_{\rm c} \approx\ 5.1 \pm0.4$, lies slightly above the nearly universal ratio of 4.3 estimated for 122-compounds (solid line) \cite{Park10}, but is close to that in FeTe$_{1-x}$Se$_x$, LiFeAs, La-1111, and cuprate superconductors.

Another important dimensionless parameter that allows an assessment of the pairing strength in unconventional superconductors is the $\hslash\omega_{\rm res}/2\Delta$ ratio, where $\Delta$ is the superconducting gap. We have recently shown that in Fe-based superconductors this ratio varies between the universal value of $\sim$\,0.64, typical for strong-coupling superconductors with high $T_{\rm c}$ \cite{Yu09}, and unity for low-$T_{\rm c}$ compounds \cite{Inosov11}. In the 245-systems, the SC gap has been measured by ARPES and NMR \cite{Gap,Ma11}, producing the average $2\Delta/k_{\rm B}T_{\rm c}$ ratio of $\sim$\,$7.2 \pm 0.4$ \cite{Inosov11}. It corresponds to the $\hslash\omega_{\rm res}/2\Delta$ ratio of $\sim$\,$0.7 \pm 0.1$ in RFS, slightly above the strong-coupling limit and supporting the general trend found in all Fe-based superconductors \cite{Inosov11}.

To conclude, we have observed the magnetic resonant mode coexisting with the AFM order in RFS below the SC transition. Our finding suggests unconventional pairing with a sign-changing order parameter in the 245-systems, qualitatively consistent with theoretical predictions, made under the assumption of finite electron doping in the metallic phase volume. Although this result points toward an electronic phase separation scenario \cite{Ricci11}, the resonant spectral weight is comparable to that in optimally doped 122-compounds. It evidences bulk superconductivity in our sample, consistent with the $\sim$100\% flux exclusion observed in the ZFC magnetization measurement, and implies an unusual coexistence or microscopic segregation of strong antiferromagnetism with superconductivity, which should be addressed by future theoretical work. The estimated ratios of $\hslash\omega_{\rm res}/k_{\rm B}T_{\rm c} \approx 5.1 \pm 0.4$ and $\hslash\omega_{\rm res}/2\Delta \approx 0.7 \pm 0.1$ in this compound indicate moderately strong pairing, similar to other Fe-based superconductors.

\textit{Acknowledgements.} We thank  A.\,V. Boris, A. Charnukha, D. Efremov, V. Hinkov, G. Jackeli, G. Khaliullin, and I.\,I. Mazin for helpful discussions. Y. Li thanks the Alexander von Humboldt foundation for finantial support. This work has been supported, in part, by the DFG within the Schwerpunktprogramm 1458, under Grants No. BO3537/1-1 and DE1762/1-1, and via TRR80 (Augsburg-Munich).


\begin{thebibliography}{10}

\vspace{-1em}
{\item[\hspace{1.7em}\large *]\hspace{-0.38em}Corresponding author: \href{mailto:d.inosov@fkf.mpg.de}{d.inosov@fkf.mpg.de}\label{CorrAuthor}\vspace{0.8em}}


\bibitem{Discovery} J.\,-G.\,Guo \textit{et~al.}, \href{http://link.aps.org/abstract/PRB/v82/e180520}{Phys.~Rev.~B {\bf 82}, 180520(R) (2010)};
                    A.\,F.\,Wang \textit{et~al.}, \href{http://link.aps.org/abstract/PRB/v83/e060512}{\textit{ibid}. {\bf 83}, 060512(R) (2011)};
                    A.\,Krzton-Maziopa \textit{et~al.}, \href{http://iopscience.iop.org/0953-8984/23/5/052203}{J.~Phys.:~Condens.\,Matter {\bf 23}, 052203 (2011)}.

\bibitem{Dispute} I.\,I.\,Mazin, \href{http://physics.aps.org/articles/v4/26}{Physics {\bf 4}, 26 (2011)};
                  Z.\,Shermadini \textit{et al.}, \href{http://prl.aps.org/abstract/PRL/v106/i11/e117602}{Phys.~Rev.~Lett. {\bf 106}, 117602 (2011)}.

\bibitem{Chen11} F.\,Chen \textit{et al.}, \href{http://arxiv.org/abs/1106.3026}{arXiv:1106.3026} (unpublished).

\bibitem{Mazin08} I.\,I.\,Mazin, D.\,J.\,Singh, M.\,D.\,Johannes, and M.\,H.\,Du, \href{http://prl.aps.org/abstract/PRL/v101/i11/e057003}{Phys.~Rev.~Lett. {\bf 101}, 057003 (2008)}.

\bibitem{ARPES} T.\,Kondo \textit{et al.}, \href{http://prl.aps.org/abstract/PRL/v101/i11/e147003}{Phys.~Rev.~Lett. {\bf 101}, 147003 (2008)};
                H.\,Ding \textit{et al.}, \href{http://iopscience.iop.org/0295-5075/83/4/47001}{Europhys.~Lett. {\bf 83}, 47001 (2008)};
                K.\,Nakayama \textit{et al.}, \href{http://iopscience.iop.org/0295-5075/85/6/67002}{\textit{ibid}. {\bf 85}, 67002 (2009)};
                D.\,V.~Evtushinsky \textit{et al.}, \href{http://link.aps.org/abstract/PRB/v79/e054517}{Phys.~Rev.~B {\bf 79}, 054517 (2009)};
                K.\,Terashima \textit{et al.}, \href{http://www.pnas.org/content/106/18/7330.abstract}{Proc.~Natl.~Acad.~Sci.~USA {\bf 106}, 7330 (2009)};
                K.\,Nakayama \textit{et al.}, \href{http://prl.aps.org/abstract/PRL/v105/i11/e197001}{Phys.~Rev.~Lett. {\bf 105}, 197001 (2010)}.

\bibitem{Hanaguri10} T.\,Hanaguri \textit{et~al.}, \href{http://www.sciencemag.org/content/328/5977/474.abstract?sid=2c0b0b2b-b0a0-4d0b-94ad-bdd538fabb0d}{Science {\bf 328}, 474 (2010)}.

\bibitem{Resonance} A.\,D.\,Christianson \textit{et al.}, \href{http://www.nature.com/nature/journal/v456/n7224/full/nature07625.html}{Nature {\bf 456}, 930 (2008)};
                   M.\,D.\,Lumsden \textit{et al.}, \href{http://prl.aps.org/abstract/PRL/v102/i11/e107005}{Phys.~Rev.~Lett. {\bf 102}, 107005 (2009)};
                   S.\,Chi \textit{et al.}, \href{http://prl.aps.org/abstract/PRL/v102/i11/e107006}{\textit{ibid}. {\bf 102}, 107006 (2009)};
                   Y.\,Qiu \textit{et~al.}, \href{http://prl.aps.org/abstract/PRL/v103/i11/e067008}{\textit{ibid}. {\bf 103}, 067008 (2009)};
                   D.\,S.\,Inosov \textit{et al.}, \href{http://www.nature.com/nphys/journal/v6/n3/full/nphys1483.html}{Nat.~Phys. {\bf 6}, 178 (2010)};
                   M.\,Ishikado \textit{et al.}, \href{http://arxiv.org/abs/1011.3191}{arXiv:1011.3191} (to be published in Physica C);
                   A.\,E.\,Taylor \textit{et al.}, \href{http://link.aps.org/abstract/PRB/v83/e220514}{Phys.~Rev.~B {\bf 83}, 220514(R) (2011)}.

\bibitem{Shamoto10} S.\,Shamoto \textit{et~al.}, \href{http://link.aps.org/abstract/PRB/v82/e172508}{Phys.~Rev.~B {\bf 82}, 172508 (2010)}.

\bibitem{Bandstructure} X.-W.\,Yan, M.\,Gao, Z.-Y.\,Lu, and T.\,Xiang, \href{http://prb.aps.org/abstract/PRB/v84/i5/e054502}{Phys.~Rev.~B {\bf 84}, 054502 (2011)};
                        C.\,Cao and J.\,Dai, \href{http://cpl.iphy.ac.cn/qikan/epaper/zhaiyao.asp?bsid=12354}{Chin.~Phys.~Lett. {\bf 28}, 057402 (2011)};
                        I.\,R.\,Shein and A.\,L.\,Ivanovskii, \href{http://www.sciencedirect.com/science/article/pii/S037596011001635X}{Phys.~Lett.~A {\bf 375}, 1028 (2011});
                        I.\,A.\,Nekrasov and M.\,V.~Sadovskii, \href{http://www.springerlink.com/content/p968541148465474/}{JETP~Lett. {\bf 93}, 166 (2011)}.

\bibitem{Qian11} T.\,Qian \textit{et~al.}, \href{http://prl.aps.org/abstract/PRL/v106/i11/e187001}{Phys.~Rev.~Lett. {\bf 106}, 187001 (2011)}.

\bibitem{Gap} X.-P.\,Wang \textit{et al.}, \href{http://iopscience.iop.org/0295-5075/93/5/57001}{EPL {\bf 93}, 57001 (2011)};
              L.\,Zhao \textit{et al.}, \href{http://link.aps.org/abstract/PRB/v83/e140508}{Phys.~Rev.~B {\bf 83}, 140508(R) (2011)};
              D.\,Mou \textit{et al.}, \href{http://prl.aps.org/abstract/PRL/v106/i11/e107001}{Phys.~Rev.~Lett. {\bf 106}, 107001 (2011)};
              Y.\,Zhang \textit{et al.}, \href{http://www.nature.com/nmat/journal/v10/n4/full/nmat2981.html}{Nat.~Mater. {\bf 10}, 273 (2011)}.

\bibitem{Maier11} T.\,A.\,Maier, S.\,Graser, P.\,J.\,Hirschfeld, and D.\,J.\,Scalapino, \href{http://link.aps.org/abstract/PRB/v83/e100515}{Phys.~Rev.~B {\bf 83}, 100515(R) (2011)}.

\bibitem{Pairing1} F.\,Wang \textit{et al.}, \href{http://iopscience.iop.org/0295-5075/93/5/57003}{EPL {\bf 93}, 57003 (2011)};
                  T.\,Das and A.\,V.~Balatsky, \href{http://prb.aps.org/abstract/PRB/v84/i1/e014521}{Phys.~Rev.~B {\bf 84}, 014521 (2011)}.

\bibitem{Pairing2} T.\,Saito, S.\,Onari, and H.\,Kontani, \href{http://link.aps.org/abstract/PRB/v83/e140512}{Phys.~Rev.~B {\bf 83}, 140512(R) (2011)};
                   I.\,I.\,Mazin, \href{http://prb.aps.org/abstract/PRB/v84/i2/e024529}{\textit{ibid}. {\bf 84}, 024529 (2011)};
                   R.\,Yu, P.\,Goswami, Q.\,Si, P.\,Nikolic, and J.-X.\,Zhu, \href{http://arxiv.org/abs/1103.3259}{arXiv:1103.3259} (unpublished).

\bibitem{Diffraction} P.~Zavalij \textit{et al.}, \href{http://link.aps.org/abstract/PRB/v83/e132509}{Phys.~Rev.~B {\bf 83}, 132509 (2011)};
                      W.\,Bao \textit{et al.}, \href{http://cpl.iphy.ac.cn/qikan/epaper/zhaiyao.asp?bsid=12618}{Chin.~Phys.~Lett. {\bf 28}, 086104 (2011)};
                      V.\,Y.\,Pomjakushin \textit{et al.}, \href{http://link.aps.org/abstract/PRB/v83/e144410}{Phys.~Rev.~B {\bf 83}, 144410 (2011)};
                      \href{http://iopscience.iop.org/0953-8984/23/15/156003}{J.~Phys.:~Condens.\,Matter {\bf 23}, 156003 (2011)};
                      V.~Svitlyk \textit{et al.}, \href{http://arxiv.org/abs/1105.6130}{arXiv:1105.6130} (unpublished).

\bibitem{Yan11} R.\,H.\,Liu \textit{et al.}, \href{http://iopscience.iop.org/0295-5075/94/2/27008}{EPL {\bf 94}, 27008 (2011)}; Y.\,J.\,Yan \textit{et al.}, \href{http://arxiv.org/abs/1104.4941}{arXiv:1104.4941} (unpublished).

\bibitem{MagneticFS} X.-W.\,Yan, M.\,Gao, Z.-Y.\,Lu, and T.\,Xiang, \href{http://link.aps.org/abstract/PRB/v83/e233205}{Phys.~Rev.~B {\bf 83}, 233205 (2011)};
                     C.\,Cao and J.\,Dai, \href{http://prl.aps.org/abstract/PRL/v107/i5/e056401}{Phys.~Rev.~Lett. {\bf 107}, 056401 (2011)}.

\bibitem{FSrecons} W.\,-G.\,Yin, C.-H.\,Lin, and W.\,Ku, \href{http://arxiv.org/abs/1106.0881}{arXiv:1106.0881} (unpublished);
                   T.\,Das and A.\,V.~Balatsky, \href{http://arxiv.org/abs/1106.3289}{arXiv:1106.3289} (unpublished).

\bibitem{Park09} J.\,T.\,Park \textit{et al.}, \href{http://prl.aps.org/abstract/PRL/v102/i11/e117006}{Phys.~Rev.~Lett. {\bf 102}, 117006 (2009)}.

\bibitem{Ricci11} A.\,Ricci \textit{et al.}, \href{http://iopscience.iop.org/0953-2048/24/8/082002}{Supercond.~Sci.~Technol. {\bf 24}, 082002 (2011)}; \href{http://prb.aps.org/abstract/PRB/v84/i6/e060511}{Phys.~Rev.~B {\bf 84}, 060511(R) (2011)}.

\bibitem{Li11} W.\,Li \textit{et al.}, \href{http://arxiv.org/abs/1108.0069}{arXiv:1108.0069} (unpublished).

\bibitem{Charnuka11} A.\,Charnuka \textit{et al.}, \href{http://arxiv.org/abs/1108.5698}{arXiv:1108.5698} (unpublished).

\bibitem{Tsurkan11} V.\,Tsurkan \textit{et al.}, \href{http://arxiv.org/abs/1107.3932}{arXiv:1107.3932} (unpublished).

\bibitem{Park10} J.\,T.\,Park \textit{et al.}, \href{http://link.aps.org/abstract/PRB/v82/e134503}{Phys.~Rev.~B {\bf 82}, 134503 (2010)}.

\bibitem{Ye11} F.\,Ye \textit{et al.}, \href{http://arxiv.org/abs/1102.2882}{arXiv:1102.2882} (unpublished).

\bibitem{Wang11} M.\,Wang \textit{et al.}, \href{http://arxiv.org/abs/1105.4675}{arXiv:1105.4675} (unpublished).

\bibitem{Wang10} M.\,Wang \textit{et~al.}, \href{http://link.aps.org/abstract/PRB/v81/e174524}{Phys.~Rev.~B {\bf 81}, 174524 (2010)}.

\bibitem{Eschrig06} M.\,Eschrig, \href{http://www.tandfonline.com/doi/abs/10.1080/00018730600645636}{Adv.~Phys. {\bf 55}, 47 (2006)}.

\bibitem{Yu09} G.\,Yu \textit{et al.}, \href{http://www.nature.com/nphys/journal/v5/n12/full/nphys1426.html}{Nat.~Phys. {\bf 5}, 873 (2009)}.

\bibitem{Inosov11} D.\,S.\,Inosov \textit{et al.}, \href{http://link.aps.org/abstract/PRB/v83/e214520}{Phys.~Rev.~B {\bf 83}, 214520 (2011)}.

\bibitem{Ma11} W.\,Yu \textit{et al.}, \href{http://prl.aps.org/abstract/PRL/v106/i11/e197001}{Phys.~Rev.~Lett. {\bf 106}, 197001 (2011)}.

\end{thebibliography}
\end{document}